# Improved Multiuser Detection in Asynchronous Flat-Fading Non-Gaussian Channels

K. Vidyullatha, *Student Member, IEEE,*
Electronics and Communication Engineering Department,
Kakatiya Institute of Technology and Science,
Warangal, Andhra Pradesh, India – 506 105.
e-mail: v_kanchanapally@yahoo.com

V. Harish, *Student Member, IEEE,*
Electronics and Communication Engineering Department,
Kakatiya Institute of Technology and Science, Warangal,
Andhra Pradesh, India – 506 015.
e-mail: harishvictory@gmail.com

S. V. N. L. Tejaswi,
Electronics and Communication Engineering Department,
Kakatiya Institute of Technology and Science, Warangal,
Andhra Pradesh, India – 506 015.
e-mail: svnlchunduriece@gmail.com

T. Anil Kumar, *Member, IEEE*
Electronics and Communication Engineering Department,
RRS College of Engineering and Technology, Patancheru,
Andhra Pradesh, India – 502 300.
e-mail: tvakumar2000@yahoo.co.in

*Abstract*— **In this paper, a new M-estimator based multiuser detection in asynchronous flat-fading non-Gaussian CDMA channels is considered. A new closed-form expression is derived for the characteristic function of the multiple-access interference signals. Simulation results are provided to prove the effectiveness of the derived bit-error probabilities obtained with this expression in asynchronous flat-fading non-Gaussian CDMA channels.**

## I. INTRODUCTION

Recently, the problem of robust multiuser detection in non–Gaussian channels has been addressed in the literature [1] & [2], which were developed based on the Huber and Hampel estimators. The Huber's estimator [3] can limit the effect of gross errors; however, the effect may still have to be large enough to reach an unacceptable level. The Hampel's estimator [4] avoids this problem by setting up three regions to reflect the influence of gross errors with different magnitudes. It is noticed that big gross errors greater than a given threshold will have no effect on the solution as a sharp rejection point may corrupt the estimation when observations are not carefully treated due to low redundancy. Hence, a new *M*–estimator is proposed for multiuser detection in asynchronous flat-fading non-Guassian channels in this paper. Further, a new closed-form expression is derived for computing the average bit-error rate (BER) under asynchronous transmission conditions in flat-fading non-Gaussian channels using the characteristic function method.

## II. SYSTEM MODEL

Consider the signal model of [5] in matrix notation, which can be written as

$$\mathbf{y}(i) = \mathbf{A}\boldsymbol{\theta}(i) + \mathbf{n}(i) \qquad (1)$$

where $\mathbf{y}(i) \cong [y_1(i),\ldots,y_N(i)]^T$, $\mathbf{n}(i) \cong [n_1(i),\ldots,n_N(i)]^T$, $\boldsymbol{\theta}(i) \cong (1/\sqrt{N})[b_1(i)g_1(i),\ldots,b_L(i)g_L(i)]^T$, $\mathbf{A} \equiv [\mathbf{a}_1,\mathbf{a}_2,\ldots,\mathbf{a}_L]$, $\mathbf{a}_l = [a_1^l,\ldots,a_N^l]^T$, $b_l(i)$ denote symbol stream for the *l*th user, and $g_l(i)$ is the *l*th channel fading coefficient (*N* is the processing gain and *L* is the number of active users). It is assumed that the sequence of noise samples *{n(i)}* is a sequence of independent and identically distributed (i.i.d.) complex random variables whose in-phase and quadrature components are independent non-Gaussian random variables with a common probability density function (pdf) *f*. The pdf of this noise model has the form

$$f = (1-\varepsilon)\aleph(0,\upsilon^2) + \varepsilon\aleph(0,\kappa\upsilon^2) \qquad (2)$$

with $\upsilon > 0$, $0 \leq \varepsilon \leq 1$, and $\kappa \geq 1$. Here $\aleph(0,\upsilon^2)$ represents the nominal background noise and the $\aleph(0,\kappa\upsilon^2)$ represents an impulsive component, with ε representing the probability that impulses occur.

## III. BER ANALYSIS

In this section, the characteristic function method is used to compute the average BER under asynchronous transmission conditions in flat-fading non-Gaussian channels. We first examine the statistics of each interferer. This analysis parallels that of [6], and so the presentation here is brief. We have the definition of interference from each interferer as $I_k = G_k W_k$, where $G_k$ is a zero-mean unit-variance guassian random variable and $W_k$ is defined in terms of random variables $P_k, Q_k, X_k, Y_k, S_k$ as

$$W_k = P_k S_k + Q_k(1-S_k) + X_k + Y_k(1-2S_k) \qquad (3)$$

where, $S_k$ is a uniform random variable over [0,1), $P_k$ and $Q_k$ are symmetric Bernoulli random variables, $X_k$ and $Y_k$ are the discrete random variables that represent the sums of different sets of independent symmetric Bernoulli random variables. Thus, given $W_k$, $I_k$ is a Gaussian random variable with zero-mean and conditional variance $\sigma^2_{I_k|W_k} = W_k^2$. This implies through (3) that $I_k$ given $P_k$, $Q_k$, $X_k$, $Y_k$, $S_k$, $B$ (where $B$ equals the number of chip boundaries with transitions in target user's signature waveform) is Gaussian and the conditional pdf for $I_k$ follows as

$$f_{I_k|P_k,Q_k,X_k,Y_k,S_k,B}(i_k)$$
$$= \frac{1}{\sqrt{2\pi}|P_k S_k + Q_k(1-S_k) + X_k + Y_k(1-2S_k)|}$$
$$\times \exp\left\{-\frac{i_k^2}{2[P_k S_k + Q(1-S_k) + X_k + Y_k(1-2S_k)]^2}\right\} \quad (4)$$

since $W_k$ may take negative values, a modulus operation is required in (3). Averaging over $P_k$, $Q_k$, $X_k$, $Y_k$ (which is equivalent to averaging over all interferers, spreading sequences and data sequences), yields

$$f_{I_k|S_k,B}(i_k)$$
$$= \frac{1}{4\sqrt{2\pi}} 2^{-(N-1)} \sum_{i\in A}\sum_{j\in B} \binom{A}{\frac{i+A}{2}}\binom{B}{\frac{j+B}{2}}$$
$$\times \left\{\sum_{l=1,2,3,4} \frac{1}{\sigma_l(i,j,S_k)} \exp\left\{-\frac{i_k^2}{2\sigma_l^2(i,j,S_k)}\right\}\right\} \quad (5)$$

where,

$$\sigma_1^2(i,j,S_k) = [1+i+j(1-2S_k)]^2 \quad (6)$$

$$\sigma_2^2(i,j,S_k) = [2S_k-1+i+j(1-2S_k)]^2 \quad (7)$$

$$\sigma_3^2(i,j,S_k) = [1-2S_k+i+j(1-2S_k)]^2 \quad (8)$$

$$\sigma_4^2(i,j,S_k) = [-1+i+j(1-2S_k)]^2 \quad (9)$$

note that to obtain (6), it was considered that $X_k$ and $Y_k$ given $B$ are independent. It is clear from (6) that the pdf of $I_k$ given $S_k$ and $B$ is not gaussian, though the functional form is a weighted summation of "Gaussian-like" terms. We postpone averaging $S_k$ here since they appear in the denominators of the exponential function arguments giving an intractable integral. The characteristic function of $I_k$, given $S_k$ and $B$, is

$$\Phi_{I_k|S_k,B}(\omega) = \frac{2^{-(N-1)}}{4}\sum_{i\in A}\sum_{j\in B}\binom{A}{\frac{i+A}{2}}\binom{B}{\frac{j+B}{2}}$$
$$\times\left\{\sum_{l=1,2,3,4}\exp\left\{-\tfrac{1}{2}\sigma_l^2(i,j,S_k)\omega^2\right\}\right\} \quad (10)$$

The $S_k$'s now appear in the numerators of the exponential function arguments and averaging can be carried out to give

$$\Phi_{I_k|B}(\omega) = \int_0^1 \Phi_{I_k|S_k,B}(\omega)dS_k$$
$$= \frac{2^{-(N-1)}}{4}\sum_{i\in A}\sum_{j\in B}\binom{A}{\frac{i+A}{2}}\binom{B}{\frac{j+B}{2}}$$
$$\times [J(i+1,j)+J(i,j-1)+J(i,j+1)+J(i-1,j)] \quad (11)$$

where,

$$J(i,j) = \begin{cases} \frac{\sqrt{\pi/2}}{|\omega|}\{Q(|\omega||i-j|)-Q(|\omega||i+j|)\}, & (j,\omega\neq 0) \\ \exp(-\tfrac{1}{2}i^2\omega^2), & j=0 \\ 1, & \omega=0 \end{cases} \quad (12)$$

then, using the fact that the $I_k$'s given $B$ are independent, we have the characteristic function for the total interference term $I$, given $B$, as

$$\Phi_{I|B}(\omega) = \prod_{k=2}^{K}\Phi_{I_k|B}(\omega) \quad (13)$$

Let, $\xi|B = I|B + n_1$, where $n_1$ is a Gaussian random variable representing the background noise, $I|B$ is the total other-user interference given $B$, and $\xi|B$ is the total disturbance given $B$. Since the other-user interference and background noise are independent, we have

$$\Phi_{\xi|B}(\omega) = \Phi_{I|B}(\omega)\Phi_{n_1}(\omega)$$
$$= \Phi_{n_1}(\omega) - [1-\Phi_{I|B}(\omega)]\Phi_{n_1}(\omega) \quad (14)$$

We use the Fourier inversion formula for the real integral to find the distribution function of $\xi|B$, $F_{\xi|B}(\cdot)$ which is to be used to calculate the BER, as

$$F_{\xi|B}(\xi) = \frac{1}{2} + \frac{1}{\pi}\int_0^{+\infty}\frac{\Phi_{I|B}(\omega)}{\omega}\sin(\xi\omega)d\omega \quad (15)$$

The conditional BER for our target user can be expressed, by symmetry, as

$$P_{e|A_1,B} = P\{\xi < -A_1 N\} \quad (16)$$
$$= 1 - F_\xi(A_1 N)$$
$$= \frac{1}{2} - \frac{1}{\pi}\int_0^{+\infty}\frac{\sin(A_1 N)}{\omega}\Phi_{\xi|B}(\omega)d\omega$$
$$= Q(\tfrac{A_1 N}{\sigma_{n_1}}) + \frac{1}{\pi}\int_0^{+\infty}\frac{\sin(A_1 N\omega)}{\omega}\times[1-\Phi_{I|B}(\omega)]\Phi_{n_1}(\omega)d\omega$$

Averaging over the pdf of $A_1$, and using the integral identity

$$\int_0^{+\infty} \sin(kx) x e^{-x^2/2} dx = \sqrt{\frac{\pi}{2}} k e^{-k^2/2} \quad (17)$$

we have,

$$P_{e|B} = \frac{1}{2}\left[1 - \frac{N}{\sqrt{\sigma_{n1}^2 + N^2}}\right] + \frac{N}{\sqrt{2\pi}} \int_0^{+\infty} [1 - \Phi_{I|B}(\omega)] \Phi_{n1}(\omega) \times \exp\left\{-\frac{1}{2}\omega^2 N^2\right\} d\omega \quad (18)$$

which finally yields,

$$P_{e|B} = \frac{1}{2} - \frac{N}{\sqrt{2\pi}} \int_0^{+\infty} \Phi_{I|B}(\omega) \Phi_{n1}(\omega) \exp\left\{-\frac{1}{2}\omega^2 N^2\right\} d\omega \quad (19)$$

where the characteristic function of noise is given by

$$\Phi_{n1}(\omega) = e^{-\{v^2(1-\varepsilon)^2 + k^3 v^2 \varepsilon^2\}\omega^2} \quad (20)$$

## IV. ROBUST MULTIUSER DETECTION

The basic idea of *M*–estimator based multiuser detection is to detect the symbols in (1) by first estimating the vector $\theta(i)$, and then extracting symbol estimates from these continuous estimates [1] & [2]. The required estimates of $\theta(i)$ are obtained by using *M*–estimators proposed by Huber [3]. *M*–estimators minimize a sum of function $\rho(.)$ of the residuals

$$\hat{\theta}(i) = \arg\min_{\theta(i) \in C^L} \sum_{n=1}^{N} \left\{ \rho\left(\Re\left\{y_n(i) - \sum_{k=1}^{L}[A]_{nk}\theta_k(i)\right\}\right) + \rho\left(\Im\left\{y_n(i) - \sum_{k=1}^{L}[A]_{nk}\theta_k(i)\right\}\right) \right\} \quad (21)$$

where $y_n(i)$ and $\theta_k(i)$ are the *n*th and the *k*th element of the vectors $y(i)$ and $\theta(i)$, respectively, $[A]_{nk}$ is the *n,k* th element of the matrix $A$, $\Im$ denotes imaginary part, and $\rho$ is a symmetric, positive-definite function with a unique minimum at zero, and is chosen to be less increasing than square. Suppose that $\rho$ has a derivative ($\psi = \rho'$), then the solution to (21) satisfies the implicit equation

$$\sum_{n=1}^{N}\left\{\psi\left(\Re\left\{y_n(i) - \sum_{k=1}^{L}[A]_{nk}\theta_k(i)\right\}\right) + \psi\left(\Im\left\{y_n(i) - \sum_{k=1}^{L}[A]_{nk}\theta_k(i)\right\}\right)\right\}[A]_{nk} = 0, k = 1,...,L \quad (22)$$

In an impulsive noise environment, a more efficient estimator can be obtained by considering a less sensitive function $\rho(.)$ of the residuals. Hence, the following penalty function $\rho(.)$ and the corresponding influence functions $\psi(.)$ are proposed (also see Figure 1)

$$\rho_{PROPOSED}(x) = \begin{cases} \dfrac{x^2}{2}, & \text{for } |x| \le a \\ \dfrac{a^2}{2} - a|x|, & \text{for } a < |x| \le b \\ -\dfrac{ab}{2}\exp\left(1 - \dfrac{x^2}{b^2}\right) + d, & \text{for } |x| > b \end{cases} \quad (23)$$

and

$$\psi_{PROPOSED}(x) = \begin{cases} x, & \text{for } |x| \le a \\ a\,\text{sgn}(x), & \text{for } a < |x| \le b \\ \dfrac{a}{b}x\exp\left(1 - \dfrac{x^2}{b^2}\right), & \text{for } |x| > b \end{cases} \quad (24)$$

where $d$ is a constant.

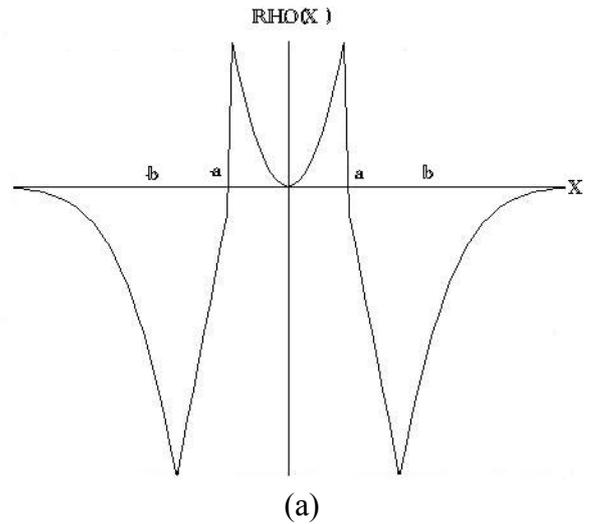

(a)

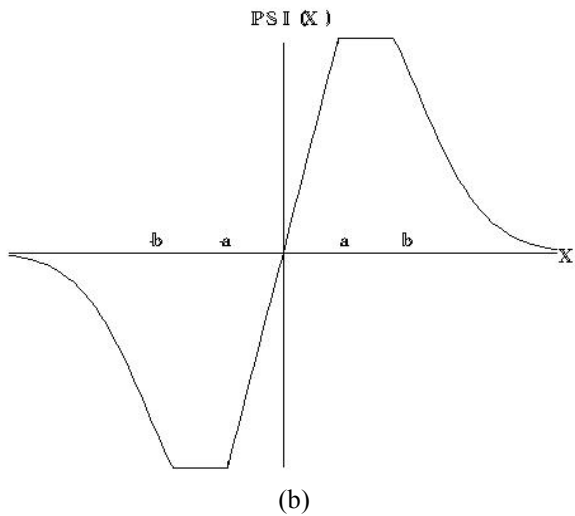

Figure 1.  (a) Penalty function and (b) influence functions of the proposed estimator

## V. SIMULATION RESULTS

In simulations, a CDMA system with 6 users, in which the spreading sequence of each user is a shifted version of *m*-sequence, is considered. The fading channel is modeled (having a perfect knowledge of the channel coefficients $g_l(i)$ ($l=1,2,..,L$)) as [5]. The performance of the proposed detector as a function of signal-to-noise ratio (SNR) in asynchronous flat-fading non-Gaussian channel (with $\varepsilon = 0.1$ and 0.01, $\kappa = 100$) for $N=127$ is shown in Fig. 2 and 3. The bit rate, the pole radius, and the spectral peak frequency have been fixed at $1/T = 10Kb/s$, $r_d = 0.998$, and $f_p = 80Hz$.

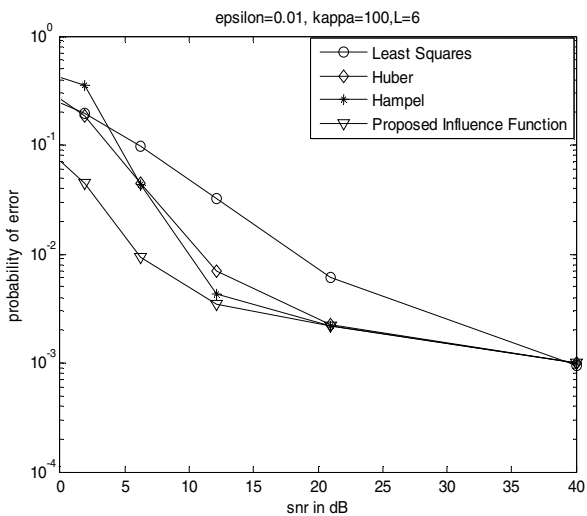

Figure 2.  Probability of error versus SNR for user 1 for the considered detectors in asynchronous flat-fading CDMA channel with non-Gaussian noise. *N =127*.

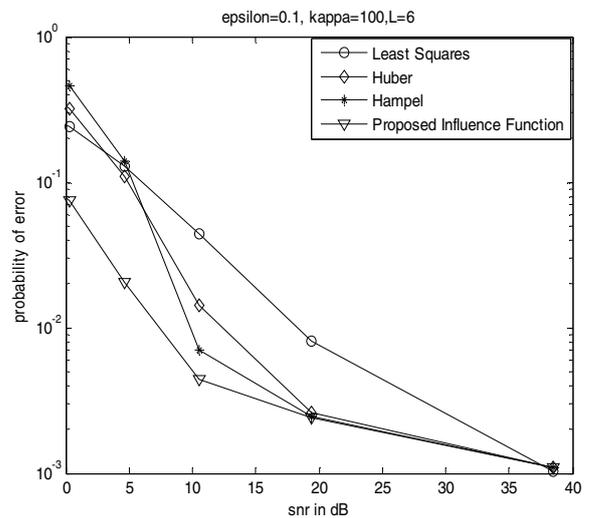

Figure 3.  Probability of error versus SNR for user 1 for the considered detectors in asynchronous flat-fading CDMA channel with non-Gaussian noise. *N =127*.

These Simulation results show that the proposed detector with the proposed influence function outperforms the linear decorrelating detector and minimax detectors (with Huber and Hampel estimators) in asynchronous flat-fading non-Gaussian channels.

## VI. CONCLUSIONS

In this paper, a new closed-form expression is derived for computing the average BER in asynchronous flat-fading non-Gaussian channels using the characteristic function method. Further a new *M*–estimator based multiuser detection technique is proposed that is seen to significantly outperform linear decorrelating detector and minimax detectors (with Huber and Hampel *M*–estimators) in asynchronous flat-fading CDMA channels with impulsive noise with little attendant increase in the computational complexity.